\documentclass[aps,prl,twocolumn,showpacs,superscriptaddress]{revtex4-2}

\usepackage{graphicx}  % Include figure files
\usepackage{amsmath}   % Advanced maths commands
\usepackage{amssymb}   % Additional math symbols
\usepackage{color}
\usepackage[T1]{fontenc} % Font encoding for international characters
\usepackage[dvipsnames]{xcolor}
\usepackage{hyperref}  % Include hyperlinks
\usepackage{acronym}
\hypersetup{
    colorlinks=true,
    linkcolor=blue,
    filecolor=blue,
    urlcolor=blue,
    citecolor=blue,
    pdfborder={0 0 0}
}

\newcommand{\SPA}{School of Physics and Astronomy, Monash University, Vic 3800, Australia}
\newcommand{\OzGravMonash}{OzGrav: The ARC Centre of Excellence for Gravitational Wave Discovery, Clayton VIC 3800, Australia}
\newacro{GW}{gravitational wave}

\begin{document}
\title{A subpopulation of low-mass, spinning black holes: signatures of dynamical assembly}

\author{Hui Tong}
\email{hui.tong@monash.edu}
\affiliation{\SPA}
\affiliation{\OzGravMonash}

\author{Thomas A. Callister}
\affiliation{Williams College, Williamstown, MA 01267, US}

\author{Maya Fishbach}
\affiliation{Canadian Institute for Theoretical Astrophysics, 60 St George St, University of Toronto, Toronto, ON M5S 3H8, Canada}
\affiliation{David A. Dunlap Department of Astronomy and Astrophysics, 50 St George St, University of Toronto, Toronto, ON M5S 3H8, Canada}
\affiliation{Department of Physics, 60 St George St, University of Toronto, Toronto, ON M5S 3H8, Canada}

\author{Eric Thrane}
\affiliation{\SPA}
\affiliation{\OzGravMonash}

\author{Fabio Antonini}
\affiliation{Gravity Exploration Institute, School of Physics and Astronomy, Cardiff University, Cardiff, CF24 3AA, UK}

\author{Simon Stevenson}
\affiliation{Centre for Astrophysics and Supercomputing, Swinburne University of Technology, Hawthorn, VIC 3122, Australia}
\affiliation{OzGrav, ARC Centre for Excellence of Gravitational Wave Discovery, Hawthorn, VIC 3122, Australia}

\author{Isobel M. Romero-Shaw}
\affiliation{Gravity Exploration Institute, School of Physics and Astronomy, Cardiff University, Cardiff, CF24 3AA, UK}

\author{Fani Dosopoulou}
\affiliation{Gravity Exploration Institute, School of Physics and Astronomy, Cardiff University, Cardiff, CF24 3AA, UK}

\begin{abstract}
Gravitational-wave observations of massive, rapidly spinning binary black holes mergers provide increasing evidence for the dynamical origin of some mergers. Previous studies have interpreted the mergers with primary mass $\gtrsim45\,M_\odot$ as being dominated by hierarchical, second-generation mergers, with rapidly spinning primaries being the products of previous black hole mergers assembled in dense stellar clusters. In this work, we reveal confident evidence of another subpopulation with rapid and isotropic spins at low mass containing the two exceptional events GW241011 and GW241110~\citep{GW241011_GW241110}, consistent with a hierarchical merger hypothesis. Our result suggests the mass distribution of the second-generation black holes is peaked at low primary masses of $\sim16\,M_\odot$ rather than $\gtrsim45\,M_\odot$ in the pair-instability gap. Such low-mass second-generation black holes must be formed from the merger of even lighter first-generation black holes, implying that dense, metal-rich stellar environments contribute to the binary black hole population. By separating the contamination of higher-generation black holes, our result reveals the primary mass distribution of first-generation black holes formed from stellar collapse, which shows a significant dip between $\sim12\,M_\odot$ to $\sim20\,M_\odot$. This may indicate a dearth of black holes due to variation in the core compactness of the progenitor.
\end{abstract}

%\linenumbers
\maketitle

{\it Introduction.}---As the fourth observing run of LIGO–Virgo–KAGRA progresses~\citep{aLIGO,aVIRGO,KAGRA:2020tym}, over 150 binary black hole (BBH) events have been announced so far. Two broad scenarios have been widely considered as plausible formation channels for these binary systems: the isolated evolution of massive binary stars, or dynamical interactions in dense stellar clusters~\citep{Mapelli_2020}. The size of the latest gravitational-wave transient catalog, GWTC-4, is now enabling meaningful exploration of and constraints on these different channels~\citep{GWTC-4_result,GWTC_4_population}. 

Several recent studies have presented compelling evidence that some BBH systems arise hierarchically in dense stellar environments~\citep{Doctor_2020,hierarchical,Kimball:2020qyd,Mould:2022ccw, Wang:2022gnx,Li_2023yyt,Antonini:2024het,Tong_gap,Antonini:2025ilj}. Hierarchical mergers are those that include at least one black hole that is itself a remnant of a previous merger~\citep{GerosaFishbach}. Such higher-generation remnants are more massive than their first-generation ancestors and are expected to have large dimensionless spins of $\chi\approx0.7$, inherited from the orbital angular momentum of their progenitor binaries~\citep{Pretorius,Buonanno,Gonzalez:2006md,Fishbach:2017dwv}. In line with this expectation, References~\cite{Li_2023yyt,Antonini:2024het,Pierra_2024, Sadiq:2025vly, Antonini:2025zzw} identified a correlation between black hole masses and spins, with systems of $m_1\gtrsim 45\,M_\odot$ tending to have a significantly broader effective inspiral spin distribution (also see References~\cite{Wang:2022gnx, Li:2025iux} which reported evidence of large but preferentially aligned spin). The effective inspiral spin, $\chi_{\rm{eff}}$, is a leading-order (i.e., typically best-measured) spin term from the post-Newtonian expansion of the GW waveform, defined as a mass-weighted sum of spin vectors projected along the angular momentum axis~\citep{Damour}:
\begin{align}
    \chi_\text{eff} = \frac{\chi_1 \cos\theta_1 + q \chi_2 \cos\theta_2}{1+q}.
\end{align}
Here, $m_1$ is the primary (more massive) black hole mass, $m_2$ is the secondary (less massive black hole mass), $\chi_{1,2}$ are the associated dimensionless spin magnitudes, and $\cos\theta_{1,2}$ are the associated cosine spin tilt angles as measured from the orbital angular momentum vector.
This observation, that the distribution of $\chi_\text{eff}$ broadens at $m_1\gtrsim45 M_\odot$ is consistent with the hierarchical merger model~\citep{Baibhav:2020xdf,Fishbach:2022lzq}. 
In addition, Ref~\citep{Antonini:2025ilj} reported marginal evidence for a spinning subpopulation at $14\,M_\odot$ although this is not statistically required by GWTC-4 data.

The existence of hierarchical mergers also provides a natural explanation for how black holes came to populate the pair-instability gap between $\approx 45-130 M_\odot$ where we expect a dearth of black holes due to (pulsational) pair instability supernovae~\citep{2016ApJ...831..187A,Gerosa:2017kvu,GerosaFishbach,Rodriguez:2019huv, Stevenson:2019rcw, Karathanasis:2022rtr, Afroz:2025ikg, Ray:2025xti}. 
Using the recently released GWTC-4 data~\citep{GWTC-4_result}, Refs.~\cite{Tong_gap,Antonini:2025ilj, Sharan_mass_ratio} have presented further evidence for a subpopulation of spinning BBH systems with high $m_1\gtrsim 45\,M_\odot$ but low $m_2\lesssim 45\,M_\odot$. 
This subpopulation can be explained by second-generation + first-generation (2G+1G) hierarchical mergers: the primary is a second-generation black hole in the pair-instability gap while the secondary formed from stellar collapse below the gap. Higher generation mergers (2G+2G, 3G+1G etc.) are expected to be much less common than 2G+1G mergers ~\citep{Rodriguez:2019huv}.

Given that the mass distribution of first-generation black holes peaks at $\approx10\,M_\odot$~\citep{Tiwari:2020otp, Tiwari:2021yvr, Tiwari:2023xff,Godfrey:2023oxb,Toubiana:2023egi,Callister:2023tgi, GWTC_4_population}, hierarchical mergers are also expected to exist at masses lower than $45M_\odot$~\citep{Ye:2025ano}. 
Two recently announced gravitational-wave events, GW241011 and GW241110, are plausible examples of low-mass 2G+1G mergers from dense star clusters~\citep{2016ApJ...831..187A,Fishbach:2017dwv, GW241011_GW241110}. Both events are unequal mass systems (mass ratio $q \approx 0.5$) with primary spin $\chi_1 \approx 0.7$. At the same time, they have distinct spin orientations. 
Whereas the primary spin of GW241011 is predominantly aligned with the orbital angular momentum (with a spin-orbit misalignment angle of $\sim 30^\circ$), the primary spin of GW241110 is likely misaligned by more than $90^\circ$.
These observations motivate the question of whether GW241011 and GW241110 belong to a distinct subpopulation of hierarchical mergers, made from previous mergers of $\approx10\,M_\odot$ black holes.

In this Letter, we analyze the newly released GWTC-4 catalog, incorporating GW241011 and GW241110, to investigate whether these two sources are consistent with belonging to a low-mass population of dynamically formed hierarchical mergers. Specifically, we ask whether they represent a subpopulation of hierarchical mergers occurring below the pair-instability gap, potentially pointing to a broader contribution from dense stellar environments to the overall binary black hole merger population.

\noindent
{\it Models.}---We analyse 155 binary black holes, consisting of 153 mergers in GWTC-4 with false-alarm rates $\le 1\,\rm{yr}^{-1}$ in addition to the two events GW241011 and GW241110 detected in the second part of the fourth observing run (O4b) of LIGO-Virgo-KAGRA. We correct for selection effects using a suite of mock signals injected into and recovered from LIGO and Virgo data~\citep{GWTC_4_injection}. 
However, we do not account for the bias that comes from the fact that, of all events detected in O4b, only GW241110 and GW241011 have been made public due to their interesting properties.
It is impossible to quantify the resulting bias---see, e.g., Ref.~\cite{leave_one_out}---so our results should be interpreted with caution until the release of the next gravitational-wave catalog containing all O4b detections.
However, we do not expect our qualitative conclusions to change, because as we show later, they are consistent with the GWTC-4 data alone.

We model the black hole mass distribution using a \textsc{Single Power Law + Two Peaks} model and allowing for a gap in the secondary mass $m_2$ distribution \cite{Tong_gap}. 
We assume that the merger rate density varies with redshift according to a power law~\citep{Fishbach:2018edt, gwtc2_pop}. 
We model the spin distribution with two approaches: a weakly modeled approach and a strongly modeled approach. 
We present the strongly modeled approach first since it helps to showcase the population features we are looking for.

Our strongly modeled approach builds on Ref.~\cite{Antonini:2024het}. 
We model the $\chi_\text{eff}$ distribution for 1G+1G mergers with a relatively narrow truncated Gaussian, which enforces our prior belief that 1G black holes tend to have modest spins.
We model the distribution for 2G+1G mergers with a broader uniform distribution, which reflects the fact that 2G black holes inherit significant spin from their progenitor binary~\footnote{The uniform distribution of $\chi_\text{eff}$ for 2G+1G can be understood as follows. 
We make several assumptions about 2G+1G binaries: that the mass ratio is typically $q\approx0.5$; the secondary spin is small so that $\chi_2\approx 0$ and the primary spin is always $\chi_1 \approx 0.7$. 
It follows that $\chi_\text{eff}$ is linearly related to the cosine of the primary tilt angle $z_1$. Since $z_1$ is approximately uniformly distributed, $\chi_\text{eff}$ is approximately uniformly distributed\citep{Antonini:2024het}.}.

\begin{figure}
    \centering
    \includegraphics[width=1.0\linewidth]{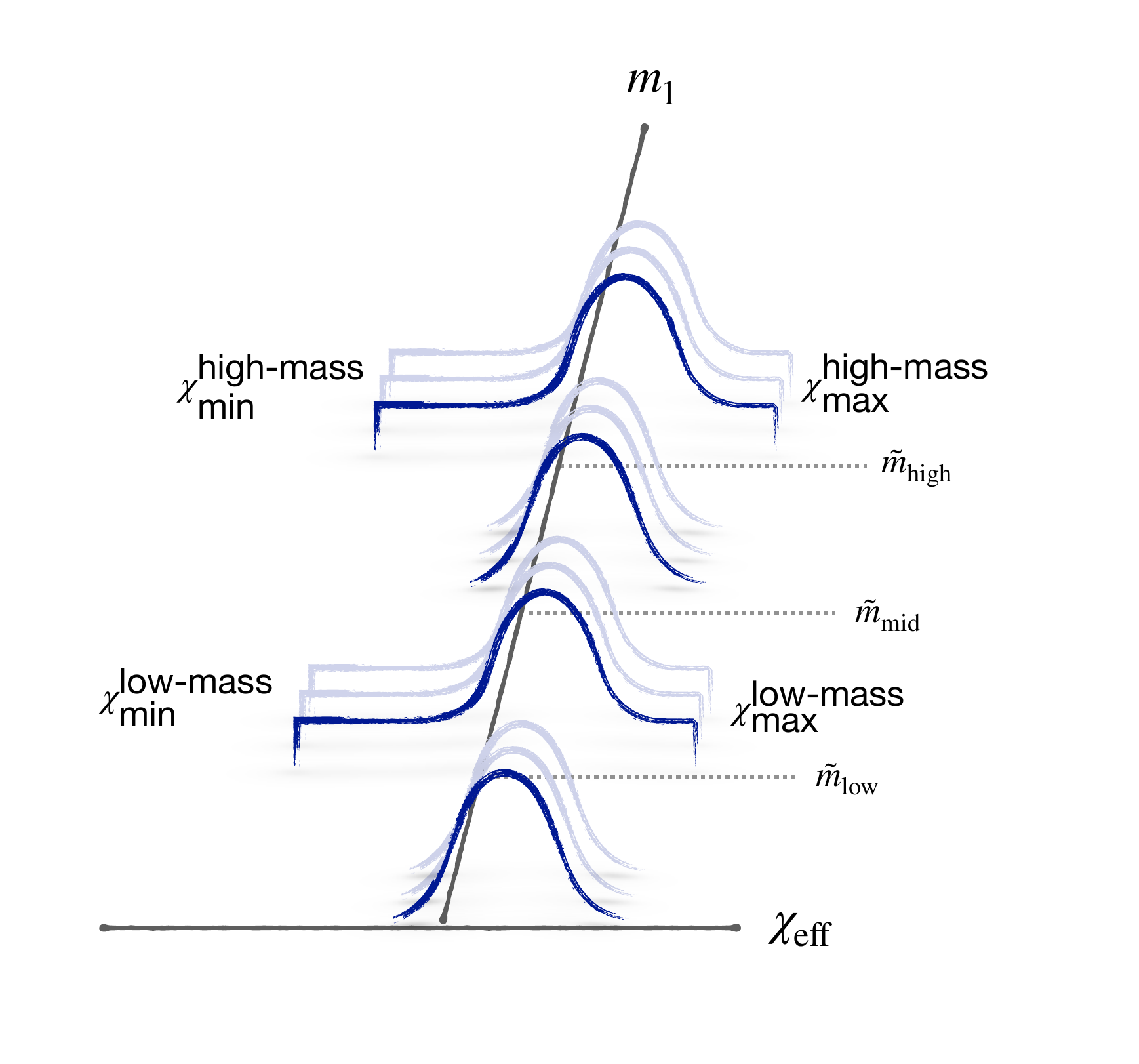}
    \caption{
    A graphical illustration of our strongly-parameterized model showing how the distribution of effective inspiral spin $\chi_\text{eff}$ varies with primary mass $m_1$.
    We interpret the truncated Gaussian distribution as arising from 1G+1G mergers while we attribute the uniform distribution to 2G+1G mergers.
    }
    \label{fig:cartoon}
\end{figure}

The model is illustrated graphically in Fig.~\ref{fig:cartoon}, which shows how the distribution of $\chi_\text{eff}$ changes with $m_1$.
We delineate four distinct regions separated by three transition masses: $\tilde{m}_\text{low}$, $\tilde{m}_\text{mid}$, and $\tilde{m}_\text{high}$:
\begin{itemize}
    \item For $m_1 < \tilde{m}_\text{low}$, the distribution of $\chi_\text{eff}$ is described with a truncated Gaussian from 1G+1G mergers.
    %%%
    \item When $\tilde{m}_\text{low} < m_1 < \tilde{m}_\text{mid}$, the distribution of $\chi_\text{eff}$ is a mixture of a truncated Gaussian from 1G+1G mergers and a uniform distribution from 2G+1G mergers.
    %%%
    \item When $\tilde{m}_\text{mid} < m_1 < \tilde{m}_\text{high}$, the distribution of $\chi_\text{eff}$ is again a truncated Gaussian from 1G+1G mergers.
    %%%
    \item When $m_1 > \tilde{m}_\text{high}$, the distribution of $\chi_\text{eff}$ is again a mixture of truncated Gaussian from 1G+1G mergers and a uniform distribution from 2G+1G mergers.
\end{itemize}
A mathematical definition of the model is provided in the Appendix; see Eq.~\ref{eq:parametric_spin}.

In all four regions the Gaussian distribution is controlled by the same two shape parameters: the mean $\mu$ and the width $\sigma$.
However, the two uniform distributions may be different.
The \textit{low-mass} distribution ($\tilde{m}_\text{low} < m_1 < \tilde{m}_\text{mid}$) is uniform on the interval $(\chi_\text{min}^\text{low-mass}, \chi_\text{max}^\text{low-mass})$.
In contrast, the \textit{high-mass} distribution ($m_1 > \tilde{m}_\text{high}$) is uniform on the interval $(\chi_\text{min}^\text{high-mass}, \chi_\text{max}^\text{high-mass})$.
The model includes two mixing fractions that control the relative contribution of the uniform distribution.
The variable $\xi_\text{low-mass}$ is the fraction of 2G+1G mergers when $\tilde{m}_\text{low} < m_1 < \tilde{m}_\text{mid}$ while $\xi_\text{high-mass}$ is the fraction of 2G+1G mergers when $m_1 > \tilde{m}_\text{high}$.
We refer to $\xi$ as the \textit{strongly modeled 2G+1G fraction}.

In the weakly modeled approach, adopted from Ref.~\cite{Antonini:2025zzw}, we again employ a mixing model with a truncated Gaussian for 1G+1G mergers and uniform distribution for 2G+1G mergers; see Eq.~\ref{eq:nonparametric_spin} in the Appendix.
This time, we fix the parameters of the uniform distribution to match expectations from 2G+1G mergers; it is symmetric around $\chi_\text{eff}=0$ with a half-width $w=0.47$~\cite{Antonini:2024het}.
We again employ a mixture fraction $f$---analogous to $\xi$ in our strong model---which determines the fraction of binaries drawn from the uniform isotropic distribution.
For this model, however, the mixing fraction is allowed to smoothly vary with primary mass $m_1$ using a Gaussian process framework developed in \cite{Antonini:2025zzw}.
We refer to $f$ as the \textit{weakly modeled 2G+1G fraction}~\footnote{We do not explicitly model the primary and secondary spin distributions, which are comparatively difficult to measure, and (in essence) assume that the only information about black hole spin comes from $\chi_\text{eff}$. We propose to revisit this assumption in future work.}.
Additional details are shown in the Supplementary Material.

\begin{figure}
    \centering
    \includegraphics[width=1.0\linewidth]{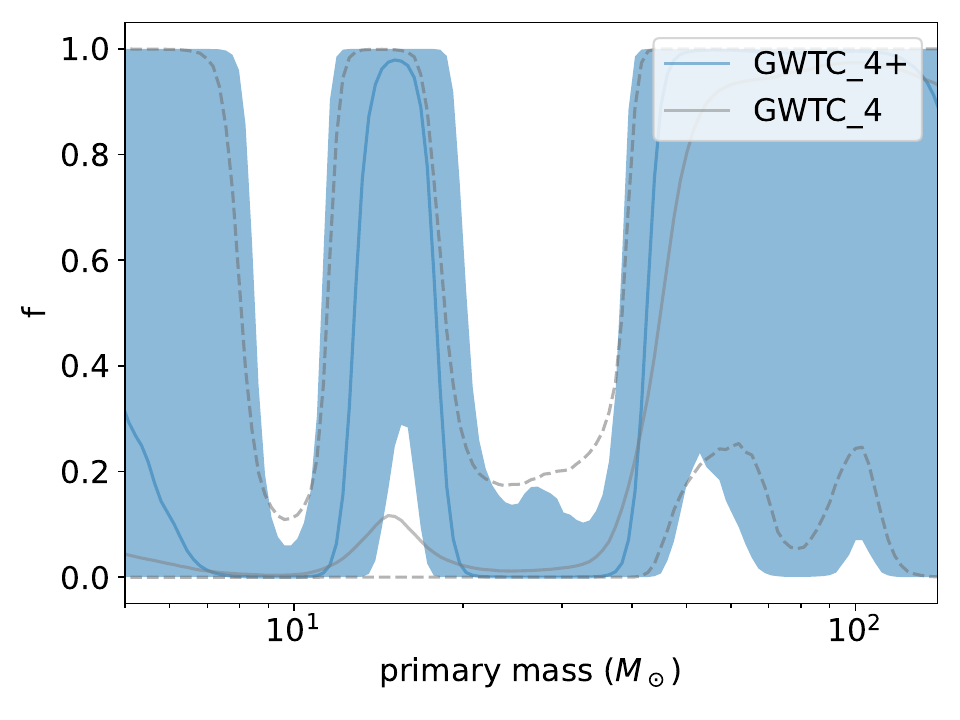}
    \caption{The weakly-modeled 2G+1G fraction as a function of primary mass. 
    The result in gray uses 153 BBH detections in GWTC-4 while blue includes GW241011 and GW241110. The median posteriors and 90\% credible intervals are presented. The fraction is consistent with zero for $m_1\sim90M_\odot$ and $m_1>120M_\odot$ since the lack of the detections makes the data uninformative in these mass scales.}
    \label{fig:fraction_GWTC_4_plus}
\end{figure}

{\it Results for the weakly modeled approach.}---In Fig.~\ref{fig:fraction_GWTC_4_plus}, we show $f(m_1)$: the weakly-modeled 2G+1G fraction as a function of the primary mass using the weakly modeled approach.
The result using GWTC-4 data is shown in gray while the result in blue includes the exceptional events GW241011 and GW241110. 
The solid curves indicate the median and the shaded bands (dashed curves) indicate the 90\% credible interval.

We reproduce a result from Ref.~\cite{Antonini:2025zzw, Antonini:2025ilj}: at $m_1 \sim 45M_\odot$, the mixing fraction transitions from $f\approx0$ to $f>0$.
That is, below $\sim45 M_\odot$, the $\chi_\text{eff}$ distribution is best described with the 1G+1G truncated Gaussian distribution while above $45 M_\odot$ it is best described with the 2G+1G uniform distribution.
The $\sim45 M_\odot$ transition has been interpreted as the lower edge of the pair-instability gap.
In this interpretation, the shift to $f\approx 1$ above $45 M_\odot$ can be understood as contamination of the gap from 2G+1G mergers~\citep{Antonini:2024het,Tong_gap,Antonini:2025ilj}.

With the inclusion of GW241011 and GW241110, we are now able to report a second 
feature at $m_1 \sim 16 M_\odot$. 
Shown in gray, this feature appeared with modest significance in GWTC-4 data although the lower limit of the fraction is consistent with zero~\citep{Antonini:2025ilj}. 
This is the mass scale at which the two exceptional events GW241011 and GW241110 sit~\citep{GW241011_GW241110}.
Including the two exceptional events, it now appears that mergers at this mass scale may come predominantly from a sub-population of 2G+1G mergers. 
At the very least, the high-spin sub-population is important near $m_1 = 16 M_\odot$ because we confidently exclude $f=0$ in this range \footnote{Sharp-eyed readers may be surprised to see that our reconstruction of the spin distribution above $m_1 \sim 45 M_\odot $ is affected by the addition of two events with primary masses that are confidently below $45 M_\odot$
The sharpness of the low-mass feature indirectly affects our inferences about the high-mass feature because both features are described by the same Gaussian process hyper-parameters.
The corner plot of the GP parameters using different datasets can be found in Fig.~\ref{fig:GWTC_4_plus_GP_parameters} in the Supplementary Materials.}.

There are two 1G+1G regions where the $\chi_{\rm{eff}}$ distribution is consistent with a truncated Gaussian distribution:  $m_1\lesssim12M_\odot$ and on the interval $m_1\sim(20M_\odot,45M_\odot)$.
We find that the truncated Gaussian is relatively narrow as expected for 1G+1G mergers $\sigma = 0.07^{+0.01}_{-0.01}$ (this and subsequent credible intervals are 90\% credibility).
Meanwhile, we find that $\mu=0.03^{+0.02}_{-0.02}$ with $\mu>0$ at 99.9\% credibility, which implies an excess of mergers with black hole spins preferentially aligned to the orbital angular momentum. 

\begin{figure*}
    \centering
    \includegraphics[width=1.0\linewidth]{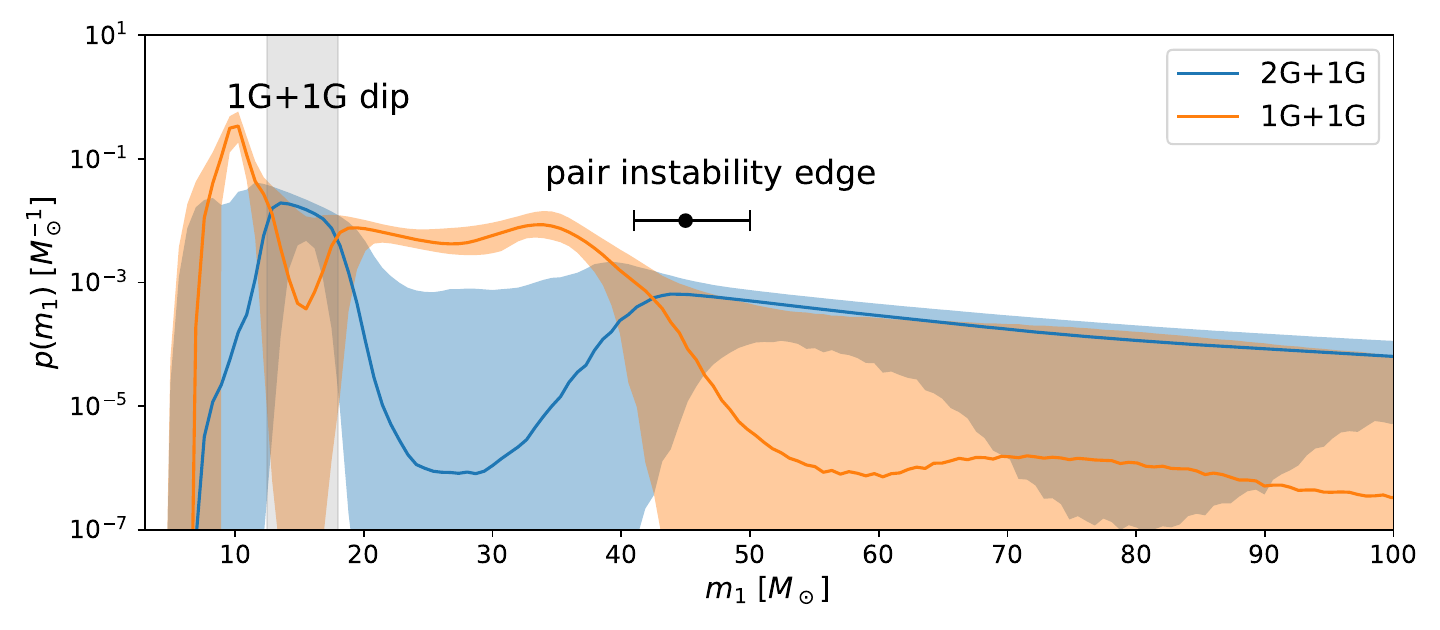}
    \caption{The primary mass distribution of binary black hole systems for 2G+1G mergers (blue) and 1G+1G mergers (orange). 
    The 1G+1G rate falls off sharply near $45 M_\odot$, which we interpret as the edge of the pair instability gap.
    We also highlight an interesting dip in the orange 1G+1G distribution between $\sim12M_\odot-18M_\odot$.
    }
    \label{fig:mass}
\end{figure*}

Figure~\ref{fig:mass} shows the reconstructed primary mass distribution for two different sub-populations.
The 1G+1G population, where $\chi_\text{eff}$ is modeled with a truncated Gaussian, is shown in orange.
The 2G+1G population, where $\chi_\text{eff}$ is modeled with a uniform distribution, is shown in blue.
The drop in 1G+1G mergers on the interval $(\sim12\,M_\odot, 20\,M_\odot)$ is consistent with the prediction from Ref.~\cite{Schneider:2020vvh}, which suggests there should be a dearth of 1G+1G black holes in this range due to the core compactness of the progenitor.
Stars might be expected to collapse and form black holes in this range, but instead experience supernovae leaving behind neutron stars due to the core compactness dip over a range of carbon-oxygen core masses.

Interestingly, our result suggests that the primary mass distribution of 2G+1G mergers may have a global peak around $16\,M_\odot$. We also note a dip in the blue 2G+1G around $30\,M_\odot$.
The over-density $\sim16M_\odot$ and under-density $\sim30\,M_\odot$ of the primary mass distribution of 2G+1G binary black holes align well with the possible mass scales of the remnants of 1G+1G mergers from the turn-on of the power-law continuum $\sim8\,M_\odot$ and dip $\sim15\,M_\odot$ of the 1G+1G binary black holes in low mass.
The primary mass distribution of 1G+1G appears to flatten out over the interval $(25\,M_\odot, 45\,M_\odot)$ before dropping sharply---consistent expectations if this is, indeed, the edge of the pair-instability gap. 
Integrating over the full range of masses, we estimate the 2G+1G merger rate density to be $2.2^{+1.9}_{-1.2}\,\rm{Gpc}^{-3}\,\rm{yr}^{-1}$.

{\it Results for the strongly modeled approach.}
Corner plots showing the posterior distributions of our strongly-modeled approach hyper-parameters are provided in the Appendix; see Figs.~\ref{fig:GWTC_4_plus_transition_mass}-\ref{fig:GWTC_4_plus_different_widths}.
The posteriors for each transition mass are highly peaked, yielding constraints consistent with our weakly-modeled reconstruction: $\tilde{m}_\text{low} = 13.1^{+2.5}_{-1.8}\,M_\odot$, $\tilde{m}_\text{mid} = 18.1^{+2.2}_{-1.7}\,M_\odot$, and $\tilde{m}_\text{high} = 42.5^{+7.0}_{-4.0}\,M_\odot$.
These results do not vary significantly if we insist that the hyper-parameters of the high-mass uniform distribution are identical to those of the low-mass uniform distribution.

Our model with three transition masses is strongly favored over a variation where we get rid of the transitions at $\tilde{m}_\text{low}$ and $\tilde{m}_\text{mid}$, with a natural log Bayes factor of $\ln\mathcal{B} = 11.7$.
It is favored over a model with no transitions with $\ln\mathcal{B} = 16.8$.

Further investigation shows that populations we identify as 2G+1G mergers have properties consistent with our expectations.
For 2G+1G mergers, we expect $\chi_\text{eff}$ to be uniformly distributed with a half-width of 0.47 \citep{Antonini:2024het}.
Both the low-mass population and the high-mass population are consistent with half-width=0.47 uniform distributions.
We constrain the half-width of the low-mass $\chi_\text{eff}$ distribution to be $0.54^{+0.21}_{-0.16}$.
For 2G+1G mergers, we expect the $\chi_\text{eff}$ distribution to be symmetric around zero so that $\langle\chi_\text{eff}\rangle=0$.
We find that $\langle\chi_\text{eff}\rangle=-0.02^{+0.16}_{-0.19}$ for both the low-mass and $\langle\chi_\text{eff}\rangle=-0.01^{+0.16}_{-0.12}$ for high-mass populations.

For the low-mass population, we infer that the strongly-modeled 2G+1G fraction is $\xi_\text{low-mass}>0.31$ at 90\% credibility.
For the high-mass population we infer that the strongly-modeled 2G+1G fraction is $\xi_\text{high-mass}>0.42$ at 90\% credibility. 
Integrating over mass, we estimate the strongly-modeled 2G+1G merger rate to be $1.08^{+1.63}_{-0.77}\,\rm{Gpc}^{-3}\,\rm{yr}^{-1}$ for the low mass spinning subpopulation and $0.26^{+0.25}_{-0.14}\,\rm{Gpc}^{-3}\,\rm{yr}^{-1}$ for the high mass spinning subpopulation.
(The sum of these two rates is consistent with the weakly-modeled rate estimated above.)

{\it Discussion and conclusions.}---We identify a subpopulation of low-mass mergers characterized by a quasi-uniform $\chi_\text{eff}$ distribution consistent with what one would expect for a 2G+1G hierarchical mergers.
This population is most prominent for mergers where the primary black-hole mass is on the interval $13.1-18.1\,M_\odot$. 
If our interpretation is correct, the 2G black holes in these binaries are necessarily formed from even lower mass first-generation black holes~\citep{GW241011_GW241110}, possibly in dense star clusters. 
Dense star clusters are formed across cosmic history at a range of metallicities, and metal-rich environments like young star clusters or nuclear star clusters are expected to produce large numbers of low-mass BBH mergers~\cite{2016ApJ...831..187A,Chatterjee:2016thb,Banerjee:2021xzp}.
Using realistic cluster formation history models from galaxy simulations coupled with $N$-body stellar dynamics, Ref.~\cite{Ye:2025ano} found that the BBH primary mass distribution from dense star clusters is expected to peak at $\sim8\,M_\odot$ for 1G+1G mergers and $\sim16\,M_\odot$ for 2G+1G mergers. The peak of 2G+1G mergers matches our results in Fig.~\ref{fig:mass}.
Other channels might also contribute to 
produce hierarchical mergers. These include formation in triples or higher multiplicity systems \cite{2021ApJ...907L..19V}, and in active galactic nuclei \cite{2019PhRvL.123r1101Y}.
For future study, it could be interesting to see the number of stellar origin black holes in clusters required to generate the low mass hierarchical mergers given the low mass 2G+1G rates we calculated. This would give some insights on the fractions of mergers from dynamical channels and other channels for the low mass binary black holes $\lesssim10\,M_\odot$.

Ref.~\cite{Schneider:2020vvh} predicted a dearth of binary black holes with component masses $\approx10-15\,M_\odot$ due to variation in the core compactness of the progenitor~\citep{OConnor:2010moj, Ugliano_2012, Sukhbold:2013yca, Ertl:2015rga, Muller:2016ujh}. 
A dip can exist in the primary mass or chirp mass distribution through the observation of LIGO-Virgo-KAGRA~\cite{Tiwari:2020otp, Tiwari:2021yvr, Schneider:2023mxe, Galaudage:2024meo, Gennari:2025nho, Tiwari:2025lit, Willcox:2025cus, Willcox:2025poh}. 
However, there is as of yet no evidence of a gap if it is imposed to both component masses~\cite{Adamcewicz:2024jkr}. The mass distribution of black holes in wide astrometric binaries discovered by Gaia also seemingly respect peaks at $\sim9M_\odot$ and $\sim33M_\odot$ and a dip between them, consistent with the primary mass distribution of 1G+1G binary black holes~\citep{Gaia_BH1, Gaia_BH2, Gaia_BH3}.

Previous studies have shown the existence of hierarchical mergers could wash out the appearance of an upper mass gap predicted by pair-instability supernova \citep{Tong_gap, Antonini:2025ilj}. 
Similarly, hierarchical mergers from less massive black hole mergers might contaminate a low-mass gap. 
Disentangling the 1G black hole distribution from 2G black hole contamination by their spin properties, our analysis shows a sharper dip than previous studies in the primary mass distribution of 1G black holes between $12\,M_\odot$ and $20\,M_\odot$.
The upcoming release of the fifth gravitational-wave catalog will help shed light on the features identified in this \textit{Letter}, and test the interpretations put forward here.

\begin{acknowledgments}
\noindent{\it Acknowledgments.} 
We thank Jack Heinzel, Vaibhav Tiwari, Vasco Gennari, Thomas Dent and Matthew Mould for helpful comments.This work is supported through Australian Research Council (ARC) Centres of Excellence CE170100004, CE230100016, Discovery Projects DP220101610 and DP230103088, and LIEF Project LE210100002.
MF acknowledges support from the Natural Sciences and Engineering Research Council of Canada (NSERC) under grant RGPIN-2023-05511, the University of Toronto Connaught Fund, and the Alfred P. Sloan Foundation. 
FA and FD are supported by the UK’s Science and Technology Facilities Council grant ST/V005618/1.
SS is supported by an ARC Discovery Early Career Researcher Award (DECRA; DE220100241).
IMRS is supported by the Science and Technology Facilities Council Ernest Rutherford Fellowship, grant number UKRI2423.
This material is based upon work supported by NSF's LIGO Laboratory which is a major facility fully funded by the National Science Foundation.
The authors are grateful for for computational resources provided by the LIGO Laboratory computing cluster at California Institute of Technology supported by National Science Foundation Grants PHY-0757058 and PHY-0823459, and the Ngarrgu Tindebeek / OzSTAR Australian national facility at Swinburne University of Technology.
LIGO was constructed by the California Institute of Technology and Massachusetts Institute of Technology with funding from the National Science Foundation and operates under cooperative agreement PHY-1764464. This paper carries LIGO Document Number LIGO-P2400105.
\end{acknowledgments}

\noindent{\it Data availability.}
Supporting code is available at Ref.~\cite{this_data_release}.

\bibliography{refs}{}

\clearpage
\appendix
\section*{Supplemental Material} 

\subsection{Basic inference setups and models}

We perform hierarchical Bayesian inference with \texttt{GWPopulation} \cite{gwpopulation, Fishbach:2018edt, 2019_Bayesian, Mandel_2019, Vitale:2020aaz}.
Our dataset includes two exceptional events GW241011 and GW241110 observed in the second part of LIGO--Virgo--KAGRA's fourth observing run and 153 binary black holes with false-alarm rates $\le 1\,\rm{yr}^{-1}$ in the cumulative Gravitational Wave Transient Catalog 4 \citep{GWTC-4_result, GWTC-4_method}. 
We use the publicly released posterior samples on Zenodo \citep{GWTC_2.1_PE_release,GWTC_3_PE_release,GWTC_4_PE_release}, generated using \texttt{Bilby} and \texttt{Dynesty} \cite{2019_bilby, 2020_bilby, dynesty}. For events in GWTC-4, we use \textsc{NRSur7dq4} samples if available \citep{Varma:2019csw}. Otherwise, the MIXED samples are used.
For events detected before O4a, we use the MIXED samples reported in GWTC-3 and GWTC-2.1 \citep{gwtc-2.1, gwtc-3}. We use standard PE samples for GW241011 and GW241110~\citep{GW241011_GW241110_data_release}. We adopt a threshold on the variance in log-likelihood estimator of 1 to manage the bias from the Monte Carlo uncertainty due to the fact of a finite number of samples from each event in hiararchical Bayesian inference~\citep{GWTC_4_population, Essick:2022ojx, Talbot:2023pex, Heinzel:2025ogf}.

Ref.~\citep{Tong_gap} has reported an evident mass gap in the secondary mass distribution which leads to an improvement in terms of Bayes factor of the mass model compared with the standard mass model widely used by literature~\citep{GWTC_4_population}. We follow Ref.~\citep{Tong_gap} to use a \textsc{Single Power Law + Two Peaks} model allowing a gap in the secondary mass $m_2$ distribution. 
The mass model is parameterized via source-frame primary mass $m_1$ and mass ratio $q=m_2/m_1$, expressed as
\begin{align}
    \pi(m_1, q | \Lambda) = \pi(m_1 | \Lambda) \, \pi(q | m_1, \Lambda),
\end{align}
which is conditioned on hyper-parameters $\Lambda$ that control the shape of the distribution.
We employ a \textsc{Single Power Law + Two Peaks} model for $\pi(m_1|\Lambda)$,
\begin{widetext}
\begin{align}
 \pi(m_1 | \Lambda)\propto & \Biggl[ \lambda_0 \mathcal{P}(m_1|\alpha, m_{1,{\rm low}}, m_{\rm high}) + \lambda_1 \mathcal{N}(m_1 | \mu_1, \sigma_1, \text{low} = m_{1,{\rm low}}, \text{high}=m_{\rm high}) \\ \nonumber
& + (1 - \lambda_0 - \lambda_1) \mathcal{N}(m_1|\mu_2, \sigma_2, \text{low} = m_{1,{\rm low}}, \text{high}=m_{\rm high}) \Biggr] S(m_1 | m_{1,{\rm low}}, \delta_{m,1}),
\end{align}
\end{widetext}
where $S$ is a tapering function ensures a smooth turn-on~\citep{GWTC_4_population}.
Our model for mass ratio allows a gap in the distribution of $m_2$:
\begin{align}\label{eq:gap_model}
    \pi(q | m_1, \Lambda) \propto 
    \begin{cases}
        0 & m_g \leq q \, m_1 \leq m_g + w_g \, , \\
        q^{\beta_q} & \text{otherwise} \, .
    \end{cases}
\end{align}
Here, $m_g$ is the lower boundary of the $m_2$ mass gap and $w_g$ is the width of the gap.
We fit the redshift evolution of mergers to a power-law distribution~\citep{Fishbach:2018edt, gwtc2_pop}.

The strongly modeled distribution $\chi_\text{eff}$ is modeled like so:
\begin{widetext}
\begin{equation}\label{eq:parametric_spin}
\pi(\chi_\mathrm{eff}|m_1, \Lambda) =
\begin{cases}
\mathcal{N}(\chi_\mathrm{eff}|\mu, \sigma) & m_1 < \tilde{m}_{\mathrm{low}} \\
\xi_{{\mathrm{low-mass}}}\,\mathcal{U}(\chi_\mathrm{eff}|\chi_\text{min}^\text{low-mass}, \chi_\text{max}^\text{low-mass}) +(1-\xi_\text{low-mass})\mathcal{N}(\chi_\mathrm{eff}|\mu, \sigma) &  \tilde{m}_{\mathrm{low}} < m_1 < \tilde{m}_\text{mid}\\
%%%
{\cal N}(\chi_\text{eff} | \mu, \sigma) & \tilde{m}_\text{mid} < m_1 < \tilde{m}_\text{high}  \\
%%%
\xi_\text{high-mass}\,\mathcal{U}(\chi_\mathrm{eff}|\chi_\text{min}^\text{high-mass}, \chi_\text{max}^\text{high-mass}) +(1-\xi_\text{high-mass})\mathcal{N}(\chi_\mathrm{eff}|\mu, \sigma) & m_1 > \tilde{m}_{\mathrm{high}}
\end{cases} .
\end{equation}
\end{widetext}
Here, $\mathcal{N}$ is a truncated Gaussian on the interval [0, 1] the mean $\mu$ and the width $\sigma$ and $\mathcal{U}$ is a uniform distribution with the minimum and maximum truncation determined by $\chi_\text{min}$ and $\chi_\text{max}$. 
The weakly modeled distribution is given by:
\begin{equation}
\begin{aligned}
\label{eq:nonparametric_spin}
\pi(\chi_\mathrm{eff}|m_1, \mu, \sigma) = 
&\big(1-f(m_1)\big) \, \mathcal{N}(\chi_\mathrm{eff}|\mu, \sigma) + \\
& f(m_1) \, \mathcal{U}(\chi_\mathrm{eff}) .
\end{aligned}
\end{equation}

The priors of all mass and redshift hyper-parameters are summarized at Table~\ref{tab:mass_redshift_priors}.
\begin{table*}
\centering
 \begin{tabular}{p{1.7cm} p{10cm} p{2.5cm}} 
 \hline
 \textbf{Parameter} & \textbf{Description} & \textbf{Prior}\\
 \hline\hline
 $\alpha$ & Spectral index of the primary mass power law & U$(-4, 12)$ \\
 $\mu_1$ & Location of the first peak & U$(5, 20)$ \\
 $\sigma_1$ & Width of the first peak & U$(0,10)$ \\
 $\mu_2$ & Location of the second peak & U$(25, 60)$ \\
 $\sigma_2$ & Width of the second peak & U$(0,10)$ \\
 $m_{\rm 1, low}$ & Lower edge of taper function & U$(3,10)$ \\
 $\delta_{\rm m,1}$ & Mass range of low mass tapering & U$(0,10)$ \\
 $\lambda_0, \lambda_1$ & Mixing fractions between power law and peaks & ${\rm Dir}(\mathbf{\alpha}=(1,1,1))$ \\
 $m_{\rm high}$ & Maximum mass for distribution & $\delta(m_{\rm high} - 300)$ \\
\tableline
$\beta_q$ & Spectral index of mass ratio power law & U$(-2,7)$ \\
$m_g$ & Lower edge of the $m_2$ gap & U$(20, 150)$ \\
$w_g$ & Width of the $m_2$ gap & U$(20,150)$ \\
$m_{\rm 2, low}$ & Lower edge of taper function in $m_2$ & U$(3,m_{\rm 1, low})$ \\
$\delta_{\rm m,2}$ & Mass range of low mass tapering in $m_2$ & U$(0,10)$ \\
\tableline
$\kappa$ & Power-law index on comoving merger rate evolution & U($-10, 10$) \\
 \hline
 \end{tabular}
 \caption{A summary of priors for mass and redshift model hyper-parameters.}
 \label{tab:mass_redshift_priors}
\end{table*}

For the spin model described at Eq.~\ref{eq:parametric_spin}, the priors are summarized at Table~\ref{tab:spin_priors}. 
\begin{table*}
\centering
 \begin{tabular}{p{1.7cm} p{10cm} p{3.5cm}} 
 \hline
 \textbf{Parameter} & \textbf{Description} & \textbf{Prior}\\
 \hline\hline
 $\mu$ & Mean of the Gaussian component & U(-1,1)\\
 $\log_{10}\sigma$ & Logarithm base 10 of the standard deviation of the Gaussian component & U$(-2, 1)$ \\
 $\tilde{m}_{\mathrm{low}}$ & the transition mass of the low mass spinning subpopulation & U(5, 30)\\
 $\tilde{m}_\text{mid}$ & the interval of the mass of the low mass spinning subpopulation & U($\tilde{m}_\text{low}$, $\tilde{m}_\text{low}$+15) \\
 $\tilde{m}_\text{high}$ & the interval of the mass of the low mass spinning subpopulation & U($\tilde{m}_\text{mid}$+5, $\tilde{m}_\text{mid}$+80)\\ 
 $\xi_\text{low-mass}$ & Mixture fraction of the low mass spinning subpopulation & U$(0, 1)$ \\
 $\xi_\text{high-mass}$ & Mixture fraction of the high mass spinning subpopulation & U$(0, 1)$ \\
 $\chi_\text{min}^\text{low-mass}$ & Minimum $\chi_{\rm{eff}}$ of the low mass spinning subpopulation  & U$(-1, 1)$ \\
 $\chi_\text{max}^\text{low-mass}$ & Maximum $\chi_{\rm{eff}}$ of the low mass spinning subpopulation & U$(-1, 1)$ \\
 $\chi_\text{min}^\text{high-mass}$ & Minimum $\chi_{\rm{eff}}$ of the high mass spinning subpopulation & U$(-1,1)$ \\
 $\chi_\text{max}^\text{low-mass}$ & Maximum $\chi_{\rm{eff}}$ of the high mass spinning subpopulation & U$(-1,1)$ \\
 \hline
 \end{tabular}
 \caption{A summary of priors for the hyper-parameters of the spin model described at Eq.~\ref{eq:parametric_spin}. Note the half-widths of the uniform components of the $\chi_{\rm{eff}}$ distribution, $\rm{w}_{{\tilde{m}_{\mathrm{low,high}}}}=(\rm{max}_{{\tilde{m}_{\mathrm{low,high}}}}-\rm{min}_{{\tilde{m}_{\mathrm{low,high}}}})/2$, are constrained to be $>0.15$ to avoid the degeneracy with the Gaussian component. This proves to be a safe choice since the measured half-widths are not railing against the priors as shown in Fig.~\ref{fig:GWTC_4_plus_different_widths}.}
 \label{tab:spin_priors}
\end{table*}

\begin{figure}
    \centering
    \includegraphics[width=1.0\linewidth]{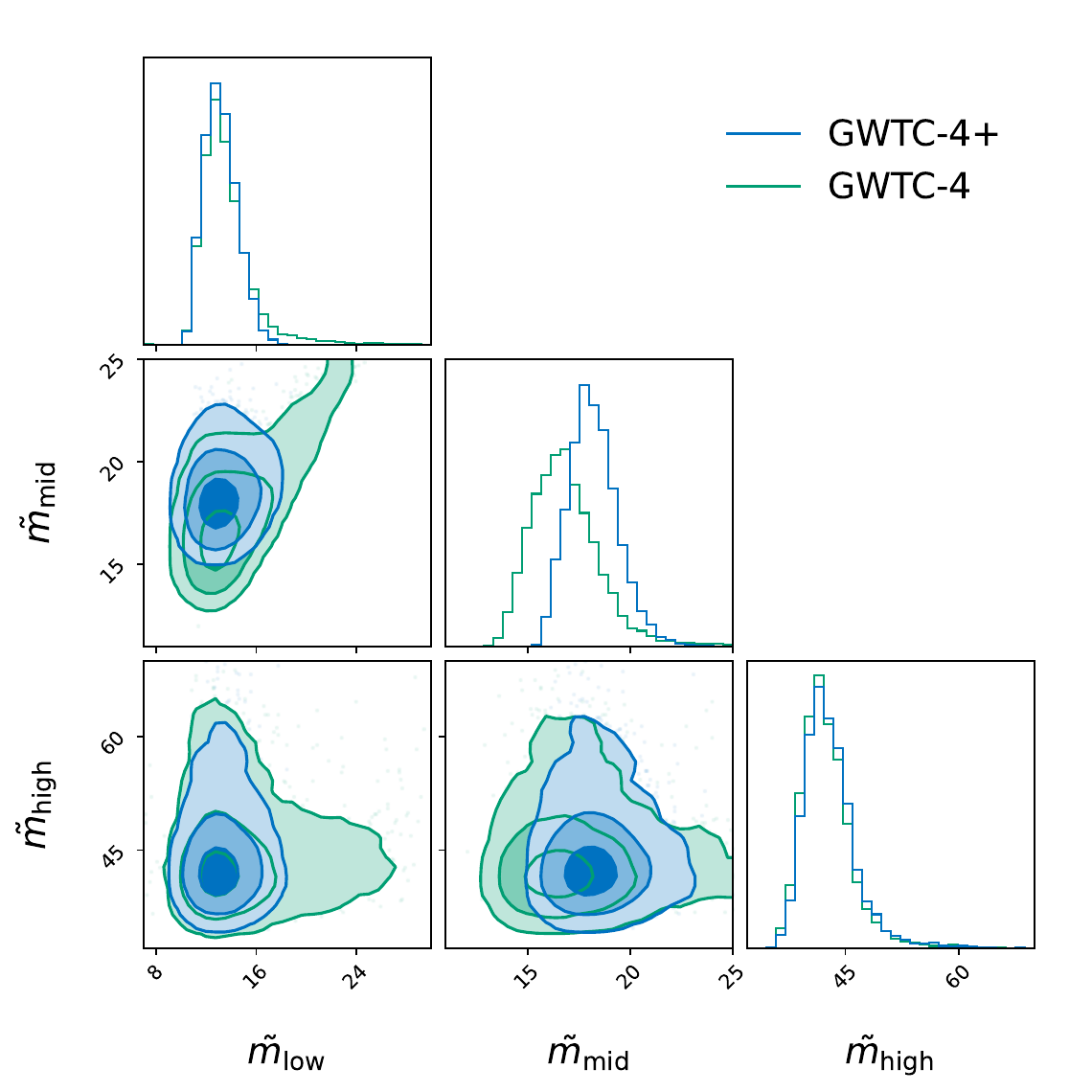}
    \caption{
    The multiple spin transition masses described in Eq.~\ref{eq:parametric_spin}. The analysis in green uses GWTC-4 events while the results in blue and orange include GW241011 and GW241110. We impose fixed mixture fractions ($\xi_\text{low-mass}=\xi_\text{high-mass}=1$) and boundaries of the uniform components ($\chi_\text{min}^\text{low-mass}=\chi_\text{min}^\text{high-mass}=-0.47$, $\chi_\text{max}^\text{low-mass}=\chi_\text{max}^\text{high-mass}=0.47$).The result from a more flexible model where the uniform distribution boundaries for low and high mass spinning subpopulations are independent and free parameters is indistinguishable from the blue trace.
    }
    \label{fig:GWTC_4_plus_transition_mass}
\end{figure}

In Fig.~\ref{fig:GWTC_4_plus_transition_mass} we show the inference on the transition masses with the strongly modeled approach, using the model allowing for two spin transitions. 
We show the results using GWTC-4 in blue and the analysis including two exceptional events in green when fixing the mixture fraction $\xi_\text{low-mass}=\xi_\text{high-mass}=1$ and assuming uniform distributions with fixed boundaries $\chi_\text{min}^\text{low-mass}=\chi_\text{min}^\text{high-mass}=-0.47$ and $\chi_\text{max}^\text{low-mass}=\chi_\text{max}^\text{high-mass}=0.47$ for the spinning subpopulations. This is the best model in our analysis. The multiple transition model is favored over single transition mass model by $\ln\mathcal{B} = 2.6$ using only GWTC-4 events.
Models with free $\xi$, $\chi_\text{min}$ and $\chi_\text{max}$ are the most disfavored by $\ln{\mathcal{B}} = -8.0$ due to Occam factor.
We relax the assumptions for either $\xi=1$ or ($\chi_\text{min}=-0.47$, $\chi_\text{max}=0.47$) respectively in the analysis to investigate the mixture fraction or the uniform distribution mean and width.

\begin{figure}
    \centering
    \includegraphics[width=1.0\linewidth]{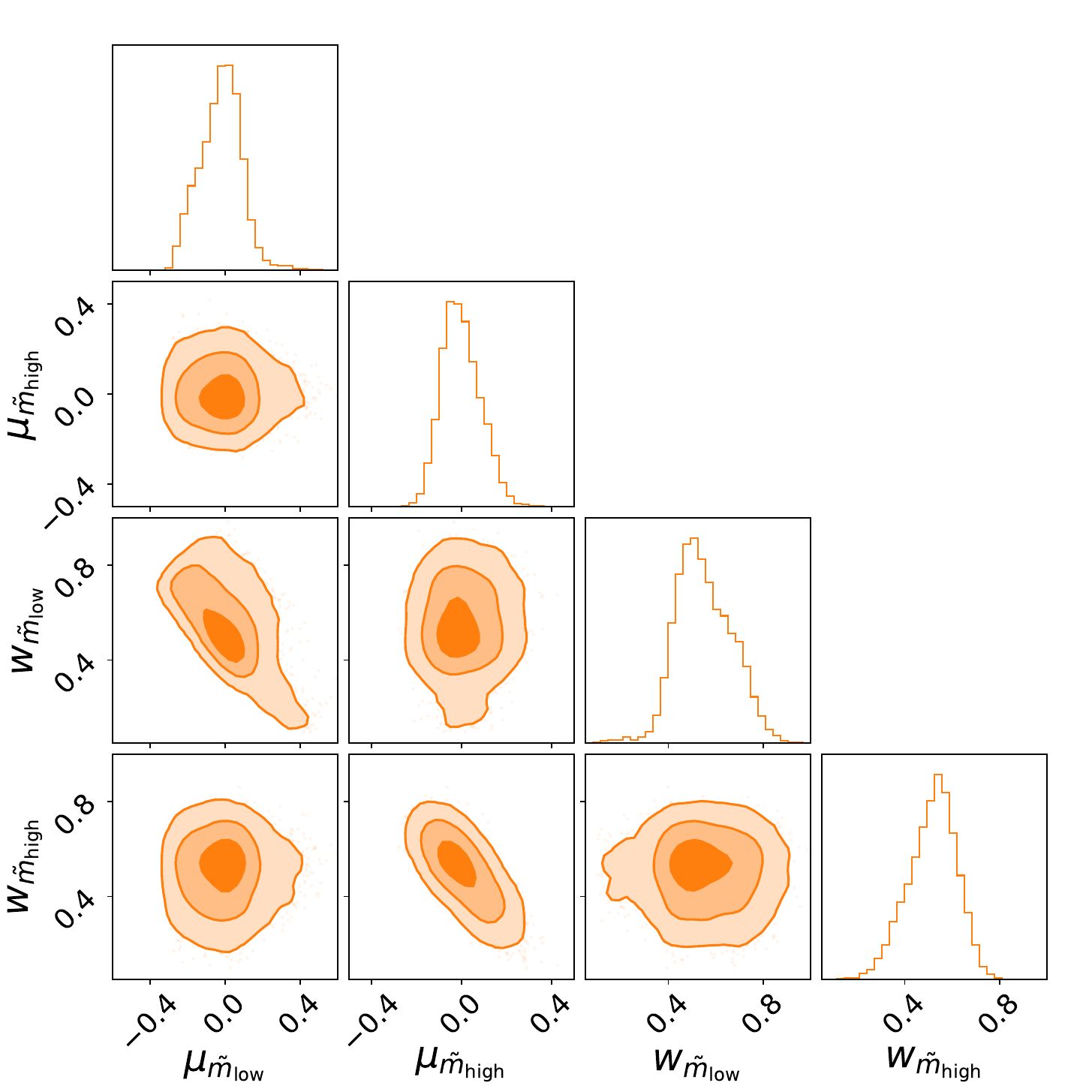}
    \caption{The measurements of the half-width and the mean of the $\chi_{\rm{eff}}$ distribution of the low and high mass spinning subpopulations in the model where the boundaries of the uniform components are free. The means are defined as $\mu_{{\tilde{m}_{\mathrm{low,high}}}}=(\rm{max}_{{\tilde{m}_{\mathrm{low,high}}}}+\rm{min}_{{\tilde{m}_{\mathrm{low,high}}}})/2$ and the half-widths are $\rm{w}_{{\tilde{m}_{\mathrm{low,high}}}}=(\rm{max}_{{\tilde{m}_{\mathrm{low,high}}}}-\rm{min}_{{\tilde{m}_{\mathrm{low,high}}}})/2$.}
    \label{fig:GWTC_4_plus_different_widths}
\end{figure}

We constrain the lower boundary of the $\chi_{\rm{eff}}$ distribution of the low mass spinning subpopulation at $\chi_\text{min}^\text{low-mass}=-0.55^{+0.27}_{-0.39}$ (90\% credibility) and upper boundary at $\chi_\text{max}^\text{low-mass}=0.51^{+0.20}_{-0.08}$ (90\% credibility). 
A positive value of minimum $\chi_{\rm{eff}}$ for the low mass spinning subpopulation is excluded at 99.1\% credibility, supporting a broad $\chi_{\rm{eff}}$ distribution with some contribution from misaligned spin binaries. 
For the high mass subpopulation, we find minimum $\chi_{\rm{eff}}<0$ at 99.7\% credibility \cite[see also Ref.][for a similar analysis]{Antonini:2025ilj}. 
There is an increasing support for mergers with negative spin in this high mass subpopulation as well since GWTC-3~\cite{Antonini:2025zzw}. 
We present in Fig.~\ref{fig:GWTC_4_plus_different_widths} the inferred means and half-widths of the uniform component of the $\chi_{\rm{eff}}$ distribution when assuming free boundaries of the uniform $\chi_{\rm{eff}}$ distributions. We calculate those parameters by defining $\mu_{{\tilde{m}_{\mathrm{low,high}}}}=(\rm{max}_{{\tilde{m}_{\mathrm{low,high}}}}+\rm{min}_{{\tilde{m}_{\mathrm{low,high}}}})/2$ and $\rm{w}_{{\tilde{m}_{\mathrm{low,high}}}}=(\rm{max}_{{\tilde{m}_{\mathrm{low,high}}}}-\rm{min}_{{\tilde{m}_{\mathrm{low,high}}}})/2$. 

\subsection{Exploring the mass ratio property of the low mass spinning subpopulation}\label{appendix:mass_ratio}
Our studies shows two subpopulations at different mass scales with spin properties consistent with hierarchical mergers.
Reference~\cite{Sharan_mass_ratio} has shown the mass ratio distribution $q$ of BBH systems above ${\sim}40\,M_\odot$ can be modeled with a Gaussian distribution peaked at $q \approx 0.5$.
This supports the interpretation of the high mass spinning subpopulation as 2G+1G hierarchical mergers.

We analyze the mass ratio distribution of this low-mass spinning subpopulation here. We fit the mass ratio distribution of the low mass spinning subpopulation to a Gaussian with free mean $\mu_q$ and width $\sigma_q$ to investigate whether the low mass subpopulation has the mass ratio distribution consistent with 2G+1G hierarchical merger prediction. The mass model is shown in Eq.~\ref{eq:mass_ratio}. We assume a Bayesian prior on $\mu_q$ that is uniform on the interval $(0, 1)$ while $\sigma_q$ is uniform on the interval $(0.1, 1)$. 
For the mergers besides this low mass spinning population, their mass ratio is modeled by a $\mathcal{P}_{\rm{gap}}(q|\beta)$, a power-law with a gap as the standard mass ratio model in Eq.~\ref{eq:gap_model}. Note we don't separately assume a different mass ratio distribution for systems with $m_1 > \tilde{m}_{\mathrm{high}}$ as their spin features. The $m_2$ gap naturally constrain the mass ratio distribution of this high mass BBH systems to be small if their secondary components are below the gap~\citep{Tong_gap} (also see Ref.~\citep{Sharan_mass_ratio} for the study on the mass ratio properties of the high mass systems).

\begin{widetext}
\begin{equation}\label{eq:mass_ratio}
\pi(q|m_1, \Lambda) =
\begin{cases}
\mathcal{P}_{\rm{gap}}(q|\beta) & m_1 < \tilde{m}_{\mathrm{low}} \\
\mathcal{N}(q|\mu_q, \sigma_q) & \tilde{m}_{\mathrm{low}} < m_1 < \tilde{m}_\text{mid}\\
\mathcal{P}_{\rm{gap}}(q|\beta) & m_1 > \tilde{m}_\text{mid}
\end{cases}
\end{equation}
\end{widetext}

\begin{figure}
    \centering
    \includegraphics[width=1.0\linewidth]{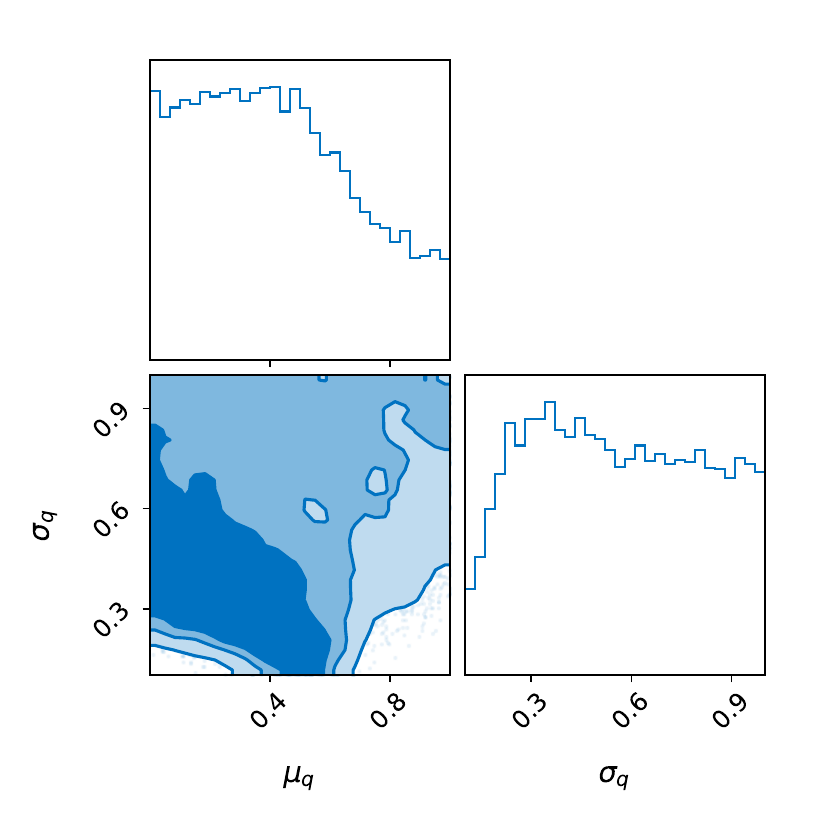}
    \caption{The mean and the width of the Gaussian distribution describing the mass ratio of the low mass spinning subpopulation. The analysis includes GW241011 and GW241110.}
    \label{fig:mass_ratio}
\end{figure}

We show the result in Fig.~\ref{fig:mass_ratio}. Although the mean $\mu_q$ is peaked at 0.5, the overall uncertainty is large and the distribution is consistent with a relatively flat mass ratio distribution with large variance of the Gaussian distribution. One interesting feature to note in the joint ($\mu_q$, $\sigma_q$) panel is, if the width of the Gaussian is forced to be small, there is stronger preference of the mean to be located around 0.5.
The mass ratio result is broadly consistent with 2G+1G hierarchical merger hypothesis. The current data is not informative enough to confidently exclude other possible formation channels from mass ratio distribution inference.

\subsection{Top candidates of the low mass spinning subpopulation}
Here we listed the binary black holes events that most likely belong to this low mass spinning subpopulation. We reported the Bayes factor for each event to come from the low mass spinning subpopulation or not using the weakly modeled approach.
GW241011 has the highest Bayes factor $\gtrsim10^7$ due to its high $\chi_{\rm{eff}}>0.42$ at 90\% credibility. Interestingly, we found the second significant event is GW231118\_005626 with Bayes factor $\sim187$. Similarly, GW231118\_005626 shows unequal mass with $q=0.55^{+0.37}_{-0.22}$ and large spins with $\chi_{\rm{eff}}=0.38^{+0.07}_{-0.08}$ and $\chi_1=0.65^{+0.28}_{-0.38}$~\citep{GWTC-4_result}. 
GW241110 is the third significant with Bayes factor $\sim61$. The significantly smaller Bayes factor compared with its twin is due to its more uncertain $\chi_{\rm{eff}}$ measurements at $-0.28^{-0.16}_{+0.16}$. GW230723\_101834 has Bayes factor $\sim 9$ with $\chi_{\rm{eff}}=-0.17^{+0.17}_{-0.13}$ and GW230706\_104333 has Bayes factor $\sim 3$ with $\chi_{\rm{eff}}=0.19^{+0.1}_{-0.1}$. The other events with primary masses falling in this region but without distinct spin features to show the preference for either subpopulation are GW230605\_065343 and GW231110\_040320.

\subsection{The impact on the high transition mass}

Our two spin transition model used in strongly modeled approach is developed from the model in \cite{Antonini:2024het} where only one spin transition is allowed. Interestingly, our analysis using the latest data shows the single transition model is not as strongly favored over no spin transition model as found in previous studies~\citep{Antonini:2024het, Antonini:2025zzw, Antonini:2025ilj}. The rapidly spinning low mass mergers impact the overall spin properties of the low mass systems. 
If not modeled properly, this low mass spinning subpopulation leads to extra ambiguity in disentangling the high mass spinning subpopulation from the low mass mergers. In this section, we discuss the impact of the lack of the flexibility of multiple spin transitions in the model. 

\begin{figure}
    \centering
    \includegraphics[width=1.0\linewidth]{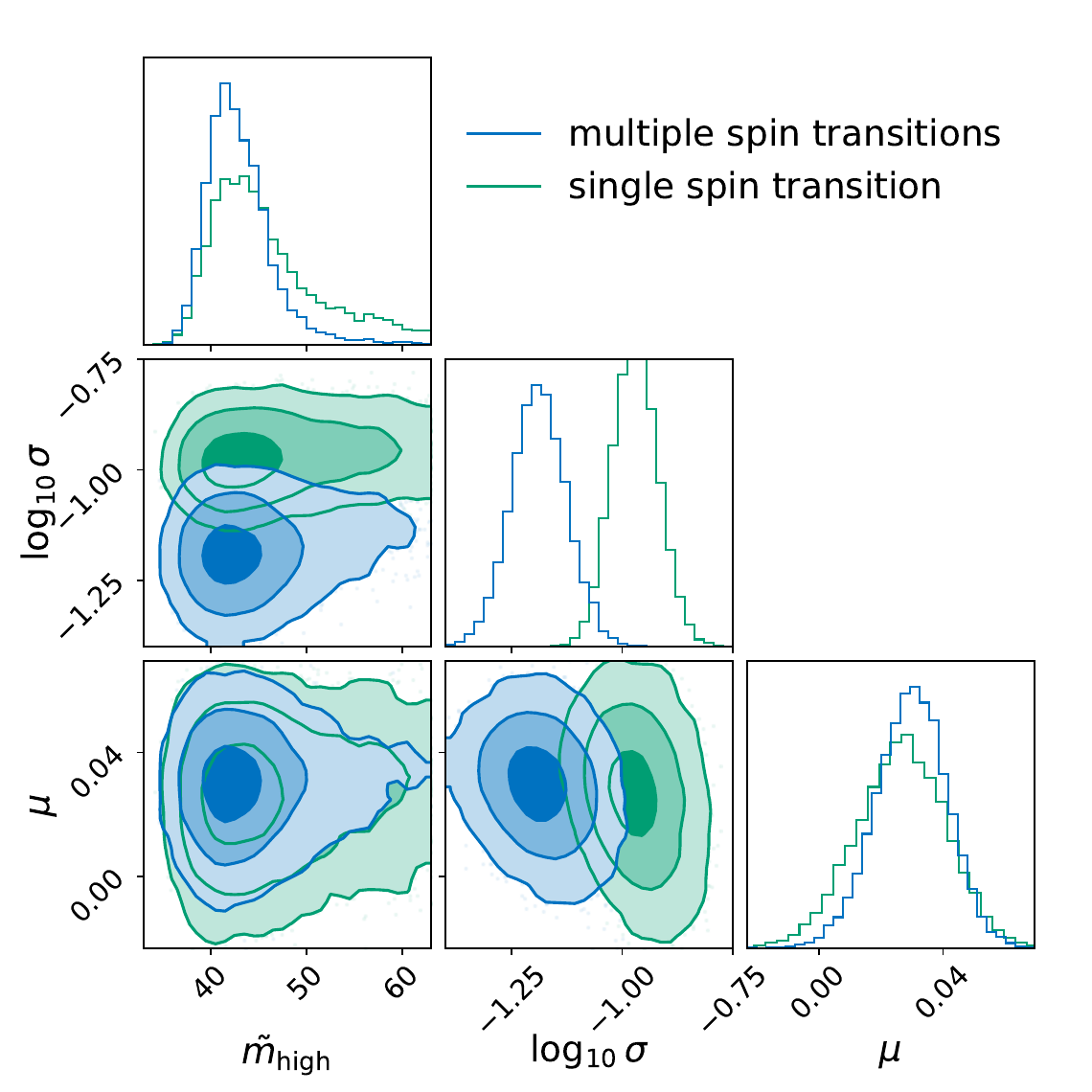}
    \caption{Comparison between the results of spin hyper-parameters using multiple spin transition model and single spin transition model. $\tilde{m}_{\mathrm{high}}$ is the transition of spin properties happened around $\sim45\,M_\odot$ which is interpreted as the mass scale where higher generation black hole mergers dominate due to lack of black holes directly from stellar collapse the pair-instability supernova \citep{Antonini:2025zzw, Tong_gap, Antonini:2025ilj}. $\mu$ and $\log_{10}\sigma$ are the mean and logarithm base 10 of the width of the Gaussian component which represents the low spin subpopulation as shown in Eq.~\ref{eq:parametric_spin}. Single spin transition model is equivalent to assume $\Delta\tilde{m}_{\mathrm{low}}=0$.}
\label{fig:GWTC_4_plus_comparision_with_single_transition}
\end{figure}

In Fig.~\ref{fig:GWTC_4_plus_comparision_with_single_transition}, we show the high mass spin transition inferred using single transition mass model and multiple spin transitions model respectively. We also present the width and mean of the Gaussian component  which represent the relatively small $\chi_{\rm{eff}}$ subpopulation.
If only one single spin transition mass is allowed, the low mass spinning subpopulation would be forced to be accounted by the Gaussian part. Although this has marginal impact of the mean of the Gaussian, the inclusion of the spinning events results in significantly broad width. Such misspecification would also lead to ambiguity of the boundary of high mass spin transition. Thus the single spin transition model will have larger uncertainty in the inference of the high mass spin transition.

\subsection{Gaussian process setup}

The mixture fraction $f(m_1)$ in the model of the weakly modeled approach is given by
\begin{equation}
    f(m_1) = S(\Phi(\ln m_1)),
\end{equation}
where $\Phi(x)$ is a draw from a zero-mean Gaussian process prior with a squared-exponential kernel
\begin{equation}
    k(x, x') = a_m^2 \exp\left(-\frac{(x - x')^2}{2\ell_m^2}\right),
\label{eq:gp-kernel}
\end{equation}
and a sigmoid function 
\begin{equation}
    S(x)=\frac{1}{1+e^{-x}}
\end{equation}
is applied in order to ensure a fraction on the bounded interval $[0, 1]$.
Within Eq.~\eqref{eq:gp-kernel}, $a_m$ and $\ell_m$ are the amplitude and length scale of the Gaussian process, respectively. We treat these as free hyper-parameters, which we fit as part of our hierarchical Bayesian inference. 

The GP prior is constructed on a uniform grid in $\log m_1$, covering the range from $2M_\odot$ to $200M_\odot$. We adopt $N_{\rm{bin}} = 200$ grid points, $\{x_i = \log m_1^{(i)}\}, \quad i = 1, \ldots, N_{\rm bin}$.
At these grid locations, the latent function values
$\mathbf{y} = [\Phi(x_1), \ldots, \Phi(x_{N_{\rm{bin}}})]^\top$ are drawn from a multivariate normal distribution, $\mathbf{y} \sim \mathcal{N}(\mathbf{0}, \mathbf{K})$, with covariance matrix $\mathbf{K}$ defined by $\mathbf{K}_{ij} = k(x_i, x_j; a_m, \ell_m)$.
To ensure numerical stability, samples from this distribution are generated using the Cholesky decomposition of $\mathbf{K}$. The resulting values of $\mathbf{y}$ are then interpolated to evaluate the GP at the positions corresponding to the posterior samples of the observed events.

\begin{figure}
    \centering
    \includegraphics[width=1.0\linewidth]{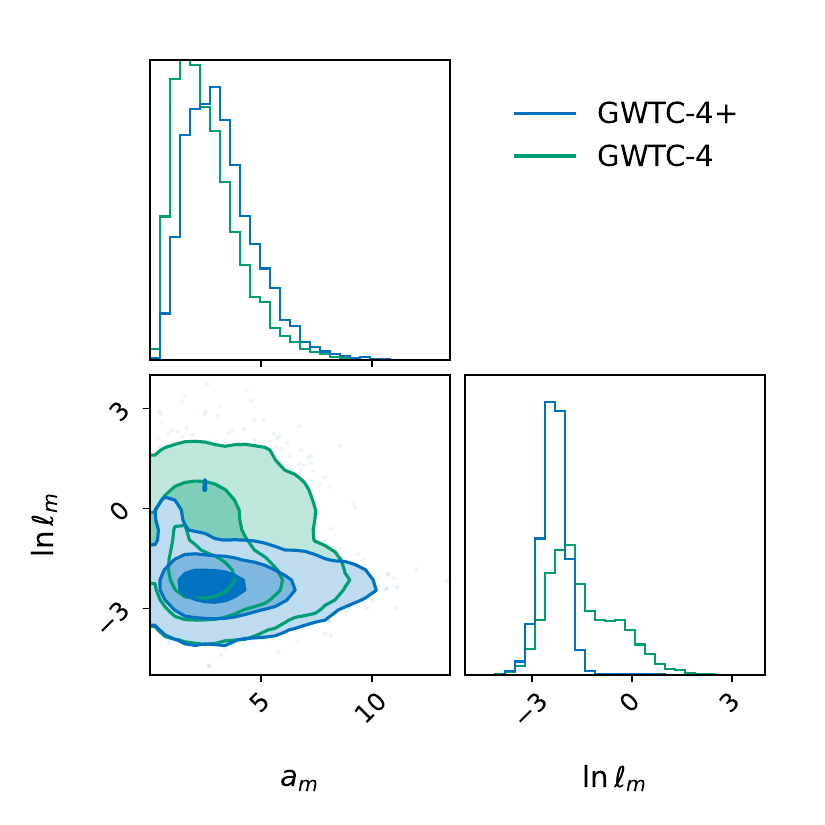}
    \caption{Corner plot of the GP hyper-parameters. Result in green uses 153 events in GWTC-4. Result in blue includes two more events GW241011 and GW241110. }
    \label{fig:GWTC_4_plus_GP_parameters}
\end{figure}

Following \cite{Antonini:2025zzw,Antonini:2025ilj}, we employ a half Gaussian with the standard deviation $=3$ as the prior for $a_m$, the amplitude of the GP, and a Gaussian with the mean $=0$ and the standard deviation $=2$ of the natural logarithm of $\ell_m$, the length scale of the GP. 
The results of the GP parameters using GWTC-4 data with and without the two exceptional events are shown in Fig.~\ref{fig:GWTC_4_plus_GP_parameters}. The inclusion of these two events constrains the $\ell_m$, the length scale of the GP prior, to be small. As discussed in the main body of this paper, this overall shorter length scale makes the GP model more sensitive to any sharp features at high mass as well. We also test other prior choices to have a half Gaussian with the standard deviation $=2$ or 1 as the prior for $a_m$. All priors produce consistent results and the half Gaussian with the standard deviation $=3$ has the highest evidence.

\end{document}